\documentclass[aps,prb,twocolumn]{revtex4}

\usepackage{graphicx}

\def\ich{$I_c(H)$}
\def\tch{$T_c(H)$}
\def\rt{$R(T)$}
\def\Ohm{\Omega}
\def\tco{T_{c0}}
\def\po{\Phi_0}

\begin{document}

\title{Enhanced vortex pinning by a composite antidot lattice in a superconducting Pb film}

\author{A.~V.~Silhanek,$^1$\footnote{present address: MST-NHMFL, MS E536, Los Alamos National Laboratory, Los Alamos, NM 87544, USA.} L. Van Look,$^2$ R. Jonckheere,$^2$ B. Y. Zhu,$^1$\footnote{National Laboratory for Superconductivity, Institute of Physics, Chinese Academy of Sciences, Beijing 100080, China.} S. Raedts,$^1$ and V.~V.~Moshchalkov$^1$}

\address{$^1$Nanoscale Superconductivity and Magnetism Group, Laboratory for Solid State Physics and Magnetism
K. U. Leuven\\ Celestijnenlaan 200 D, B-3001 Leuven, Belgium.}

\address{$^2$IMEC vzw, Kapeldreef 75, B-3001 Leuven, Belgium.}

\date{\today}

\begin{abstract}

The use of artificial defects is known to enhance the superconducting critical parameters of thin films. In the case of conventional superconductors, regular arrays of submicron holes (antidots) substantially increase the critical temperature $T_c(H)$ and critical current $I_c(H)$ for all fields. Using electrical transport measurements, we study the effect of placing an additional small antidot in the unit cell of the array. This composite antidot lattice consists of two interpenetrating antidot square arrays with a different antidot size and the same lattice period. The smaller antidots are located exactly at the centers of the cells of the array of large antidots. We show that the composite antidot lattice can trap a higher number of flux quanta per unit cell inside the antidots, compared to a reference antidot film without the additional small antidots in the center of the cells. As a consequence, the field range in which an enhanced critical current is observed is considerably expanded. Finally, the possible stable vortex lattice patterns at several matching fields are determined by molecular dynamics simulations. 
 
\end{abstract}

\pacs{PACS numbers: 74.76.Db, 74.60.Ge, 74.25.Dw, 74.60.Jg,74.25.Fy}

\maketitle

\section{Introduction}

During the last decade, compelling evidence has shown that the introduction of
an array of micro-holes (antidots) in a superconducting film has a profound
influence on both the critical current\cite{moshchalkov,metlushko} \ich~ and the critical
temperature\cite{welp,rosseel} \tch. Typically, at temperatures used for
transport measurements, the antidots are able to trap only 
one flux quantum $\Phi_0$ before saturation sets in. In this case, after the first
matching field $H_1$, interstitial vortices appear in the sample,
creating a ``composite vortex lattice", where part of the vortices
is strongly pinned at the antidots and the rest occupies
interstitial positions in between the antidots.\cite{metlushko} Due to their higher
mobility, the presence of interstitial vortices lowers the
critical current \ich~ significantly and broadens the \rt~
transition.

In this work, we study a \textit{composite antidot array},
consisting of two interpenetrating square lattices with the same
period $d=1.5~\mu$m, but different antidot size ($a_1=0.55$$\mu$m
and $a_2=0.25~\mu$m). The two sublattices are shifted with respect
to each other by half a unit cell along $x$- and $y$-directions, so that the small antidot is
placed in the center of the unit cell of the lattice of large
antidots (see Fig.~\ref{fig:ChC_layout}). This arrangement of antidots corresponds to the vortex lattice configuration at the second matching field in a sample with a single square array of antidots with $n_s=1$.  The purpose of this experiment is to
enlarge the field range where an enhanced critical current \ich~
is observed, by having efficient pinning sites exactly at the
locations where the interstitial vortices would appear if the smaller antidots were not present.

This paper is organized as follows. In the next section we present some details concerning the sample preparation and characterization. In section \ref{composite_tch} we study the phase boundary $T_c(H)$ which allows us to identify the different vortex configurations. Section \ref{section:ChC_ich} is devoted to the flux pinning properties of this composite array of holes by measuring the critical current as a function of field and temperature. We show that a composite antidot lattice considerably increases the critical current at high fields. This effect results from the fact that for $H>H_2$ the saturated small holes force the incoming vortices to occupy the big antidots. This situation persists until $H=H_4$ where the big antidots saturate and interstitial vortices form a more complex pattern. Finally, in Section \ref{patterns} we determine the most stable vortex patterns by means of molecular dynamics simulations.

\section{Sample preparation and characterization}

The used sample is a 50 nm-thick Pb film with a composite antidot lattice. The results obtained with this sample are directly contrasted with those measured on a reference antidot sample without the small holes, i.e. $a_1=0.5~\mu$m and $a_2=0~\mu$m. In both cases, the bridge made for transport measurements (see Fig.~\ref{fig:ChC_layout}(a)) has a width of 300~$\mu$m and a voltage
contact separation of 2~mm. The unit cell of the composite antidot array is shown schematically in Fig.~\ref{fig:ChC_layout}(b). The procedures followed to grow the samples are described in Ref.[\onlinecite{sophie-crete}]. The transport measurements were carried out in a commercial PPMS-Quantum Design device with a temperature stability better than 0.5 mK. All measurements were performed with the field $\bf H$ applied perpendicular to the surface of the film. The critical temperature $\tco=7.207$~K was determined from the resistive transition \rt~ in zero field, using a criterion of 10\% of the normal state resistance $R_n$.

\begin{figure}[hctb]
  \centering
  \caption{Layout of the Pb film with a composite array of square
  antidots of two different sizes. (a) Geometry of the sample showing the
  patterned area in dark gray. (b) Schematic presentation of a unit cell
  of the antidot array. (c) Atomic force micrograph of a $5 \times 5~\mu$m$^2$
  area of the composite antidot array. The lattice period
  $d$ is 1.5~$\mu$m, the antidot sizes are  $a_1=0.55~\mu$m and $a_2=0.25~\mu$m.}
  \label{fig:ChC_layout}
\end{figure}

Due to the lateral nanostructuring, the effective width of the
sample is reduced from 300~$\mu$m to 140~$\mu$m. Here, we have
assumed that we can model the antidot sample as a set of 200
parallel strips of width 0.7~$\mu$m$~(=2 \cdot 0.35~\mu$m). This
effective width was employed to calculate the resistivity
$\rho$(7.5~K)$=5.33 \cdot 10^{-8}~\Ohm$m from the
resistance $R$(7.5~K)$= 15.2~\Ohm$. Using the
listed value\cite{handbookchemistryandphysics} for $\rho \ell=4.88 \cdot 10^{-16} \Ohm $m$^2$, this resistivity value gives
an elastic mean free path of $\ell=9~$nm, and therefore a
superconducting coherence length $\xi(0)=$~25~nm (in the
dirty limit). These values are
noticeably smaller than those obtained for the reference antidot sample ($\ell=27~$nm and $\xi(0) \approx 40$~nm). Since  in a film without antidots coevaporated with the sample containing the composite antidot was obtained $\ell=27~$nm, this difference seems to be caused by the more complex lift-off procedure due to the presence of the small holes.

Knowing the mean free path $\ell$ and using the London penetration depth for the bulk\cite{lambda} Pb we obtain $\lambda(0)=71$~nm. Due to the perforation, the effective penetration depth increases, and therefore $\lambda$ should be modified according to\cite{wahl}

\begin{equation}
\Lambda_a(0)=\frac{\lambda(0)}{\sqrt{1-2
\frac{S_a}{S_t}}}=86~\mathrm{nm} \, ,
\end{equation}
where $S_a$ and $S_t$ are the area of the holes and the total area per unit cell, respectively.
As a result, the Ginzburg-Landau parameter $\kappa$ amounts to
$\kappa=\frac{\Lambda(0)}{\xi(0)}=3.4 > \frac{1}{\sqrt{2}}$, and
therefore this sample is a Type-II superconductor.\cite{comment}

The sample has been characterized by means of atomic force
microscopy. An AFM topograph of a $5 \times 5~\mu$m$^2$ area of
the film containing a composite antidot lattice is shown in
Fig.~\ref{fig:ChC_layout}(c). The root-mean-square roughness on a 1~$\mu$m$^2$
area of the sample in between the antidots is $\sigma_{RMS}=3$~nm.
This value is about two times larger than for the plane film and the reference
sample with antidots. This difference reinforces the idea that the film with the composite antidot lattice has suffered a small degradation due to a more complicated lift-off procedure.

\section{Superconducting \tch~ phase boundary}
\label{composite_tch}

\subsection{Experimental results}

We have measured the critical temperature \tch~ as a function of field for the sample with a
composite antidot lattice. The results obtained with a resistance
criterion of 10 \% of the normal state resistance $R_n$ and a
measuring current of $I_{\mathrm{ac}}=10~\mu$A are shown in
Fig.~\ref{fig:ChC_tch}, together with the phase boundary obtained for the reference antidot film. The solid line depicts the expected upper critical field boundary of a plain film with the same coherence length as the reference antidot sample according to $H_{c2}=\Phi_0/2\pi \xi(T)^2$. It is important to notice that the measured boundary of the reference antidot film is close to the $H_{c3}(T)$ line corresponding to the surface nucleation of superconductivity around the holes, whereas the solid line represents the bulk superconducting transition $H_{c2}(T)$. As a rule, the experimentally determined critical temperature \tch~ of a patterned sample turns out to be higher than that obtained for a plain film with the same coherence length.\cite{rosseel}

\begin{figure}[hctb]
  \centering
  \caption{\tch~ phase boundary for the film patterned with a composite antidot
  array, measured with an ac current of $I_{\mathrm{ac}}=10~\mu$A and a resistance
  criterion of $R_{\mathrm{crit}}=10\%~R_n$ (filled symbols). The open symbols show the phase boundary obtained for the reference antidot sample using the same criterion. The solid line is the
  calculated linear \tch~ phase boundary for a plain film with the
  same coherence length $\xi(0)=40$~nm as the antidot patterned film. The field axis is
  normalized to the first matching field $H_1=9.2$ G. The temperature axis is normalized to $T_{c0}$, the transition temperature at $H=0$.}
  \label{fig:ChC_tch}
\end{figure}

Due to the presence of the antidot array, matching features appear
in \tch~ with a periodicity of $H_1=\Phi_0/d^2=9.2$~G, corresponding to the
lattice parameter $d=1.5$~$\mu$m. Although not all of them are very
pronounced, local maxima are visible in the \tch~ of the composite array for all integer
matching fields $H_n$ ($n=1$,2, ...,6), whereas no evidence of rational
matching features is observed. Thus, \textit{the addition of the extra
antidot in the center of the unit cell of the array with large
antidots, leaves the matching period unchanged}. This is an
important observation, since the composite antidot lattice can
also be regarded as a square lattice, tilted by 45$^\circ$, with a
unit cell twice as small as that of the original lattice. If this
were the periodicity felt by the vortices, the matching period
would amount to 18.4~G, which is twice as large as the observed
period. In that case, one would expect the local maxima at even
matching fields $H_n$ ($n=2$,4, ...) in Fig.~\ref{fig:ChC_tch} to
be more pronounced than the ones at odd matching fields $H_n$
($n=1$,3, ...). Since this is not the case, we conclude that all
these peaks correspond to integer matching fields, indicating that
\textit{the main period felt by the vortices is the period of the
lattice with large antidots}.

In order to identify the vortex patterns at the matching fields,
we have plotted the $R(T)$ transition width $\Delta$\tch$=T_c(R_{\mathrm{crit}}=97 \% R_n)-T_c(R_{\mathrm{crit}}=0.1 \%
R_n)$ as a function of $H$ in Fig.~\ref{fig:ChC_deltatch} (filled symbols). In this plot three different regimes can be clearly distinguished. For $H<H_4$, the coherence length is larger than the width of the strands thus leading to a parabolic background in the $T_c(H)$ phase boundary. In this so-called ``collective" regime, we observe that the \rt~ transition width remains almost constant. For fields higher than $H_4$, an increase of the transition width can be observed, superposed with matching features at $H_5$ and $H_6$. We interpret the \textit{sudden increase in the transition width as a crossover to the regime where interstitial vortices appear in the sample}.

The interstitial regime is indicated by the gray area in Fig.~\ref{fig:ChC_deltatch} for the composite array. This
regime ranges up to 3.6~$\xi(T)=d-a$, i.e. up to $\sim H_{8}$, where a change in the \mbox{$\Delta$\tch}~ slope can be observed. For
higher fields, the single object regime is entered, where a linear phase boundary slightly distorted by an oscillation with period\cite{rosseel} $H^*=\Phi_0/a_1^2 \sim 69$ G, is expected. Although the linear phase boundary is indeed observed, single object oscillations are difficult to resolve in the narrow field range investigated. For comparison, in the same figure we show $\Delta$\tch$=T_c(R_{\mathrm{crit}}=99 \% R_n)-T_c(R_{\mathrm{crit}}=0.1 \%
R_n)$ for the reference antidot sample (open symbols). From this curve we can infer that if the smaller additional antidots are absent, the
crossover to the interstitial regime occurs at $H \sim 1.5~H_1$. Therefore the presence of the additional
smaller antidots has substantially delayed the appearance of
interstitial vortices. From the \mbox{$\Delta$\tch}~ curve, we thus conclude that \textit{the total number of trapped
flux quanta per unit cell of the antidot lattice is at least
four}. 

\begin{figure}[hctb]
\centering
\caption{Filled symbols: transition width $\Delta$ \tch$=T_c(R_{\mathrm{crit}}= 97 \% R_n)-T_c(R_{\mathrm{crit}}=0.1 \% R_n)$
of the film with a composite antidot array, measured with a current of $I_{\mathrm{ac}} =10~\mu$A. The gray box marks
the ``interstitial'' regime, where the increase of \mbox{$\Delta$\tch}~ indicates the appearance of interstitial vortices.
Open symbols: transition width $\Delta$ \tch$=T_c(R_{\mathrm{crit}}= 99 \% R_n)-T_c(R_{\mathrm{crit}}=0.1 \% R_n)$
of the reference antidot film for $I_{\mathrm{ac}} =10~\mu$A. The thin black arrows indicate the ordinate scale for each curve.}
\label{fig:ChC_deltatch}
\end{figure}

In other words, the observed saturation number is much larger than in the
reference antidot sample,
where only a $\Phi_0$-vortex can be pinned per antidot.
Even taking into account the fact that the large antidots in the
composite antidot lattice are slightly larger ($a=0.55~\mu$m) than
in the reference antidot sample ($a=0.5~\mu$m) and the difference
in coherence length, this is still a rather surprising
observation. Indeed, the addition of \textit{one} small hole per unit
cell of the antidot array leads to an unexpected \textit{increase
of the number of pinned flux quanta per unit cell from one to four}.

\subsection{Discussion}

To determine the number of flux quanta located in the large and the small antidots of the composite antidot array, a
further investigation of the \tch~ phase boundary is needed. As explained in Ref.[\onlinecite{rosseel}], the background of the phase boundary
is parabolic as long as 1.8 $\xi(T)>d-a$ (collective regime). In a
square antidot lattice, this envelope is described by\cite{pannetier}
\begin{equation}
T_c(H)=T_{c0} \left[1- \left( \frac{\mu_0 H \; \xi(0) \; \pi \;
(d-a)}{\sqrt{3} \Phi_0}\right)^2 \right] \, .
\label{parabolicbackground}
\end{equation}
By fitting the \tch~ data points at the integer matching fields
with this formula, the effective width $d-a$ of the strands between
the antidots can be deduced.

\begin{figure}[hctb]
  \centering
  \caption{\tch~ of the sample with a composite antidot array (open
  symbols). The solid line is a fit of the parabolic background (Eq. (\protect \ref{shiftedparabolicbackground}))  shifted in field by one matching period. This line nicely interpolates
  between the $T_c(H_n)$ points at the four matching fields ($n = 2, ..., 5$).}
  \label{fig:ChC_tchfit}
\end{figure}

This procedure is, of course, not {\it a priori} valid for a composite
antidot lattice. However, part of the \tch~ phase boundary
clearly shows a parabolic background. The parabola providing the
best envelope for the data points between $H_2$ and $H_4$ is
depicted by a solid line in Fig.~\ref{fig:ChC_tchfit}. For fields
higher than $H_5$, the measured curve deviates from the fitted
parabola, as is expected for the ``interstitial'' regime starting
for fields higher than $H_4$. Strikingly, the fitted
parabolic background turns out to be shifted by
one matching period, having its maximum at $H_1$ instead of at
$H=0$. By allowing this shift, the following parabolic envelope
can be found for the second up to the fourth (or even fifth)
matching peak, described by
\begin{equation}
T_c(H)=T_{c0}' \left[1- \left( \frac{\mu_0 (H-H_1) \; \xi(0) \; \pi \;
(d-a_{\mathrm{eff}})}{\sqrt{3} \Phi_0}\right)^2 \right] \, ,
\label{shiftedparabolicbackground}
\end{equation}
using $\tco '=7.195$~K and an effective antidot size of
$a_{\mathrm{eff}}=0.72~\mu$m. From this, we deduce that, after the
second matching field $H_2$, {\it the film with a composite
antidot lattice with a filled small antidot behaves as if
it would have contained a single square antidot array, but with a
larger antidot size} ($a_{\mathrm{eff}}=0.72~\mu$m). 

The following scenario might explain such a behavior. Up to $H_1$,
the vortices will be attracted towards the large antidots. Between
$H_1$ and $H_2$, vortices begin to occupy the small antidots.
Due to their size, these small antidots
trap at most a single quantum vortex. They will therefore be
completely saturated at $H_2$, creating a repulsive potential at
the position of the small antidot. Fig.~\ref{fig:ChC_potential}
shows a schematic evolution of the potential landscape along a
diagonal of the array (see dotted line in the inset) that would be experienced
by a vortex for $H=0$, $H_1$, and $H_2$. Since the large antidots pin one flux quantum,
at $H=H_1$ a surface barrier has emerged at the antidot edges. For $H=H_2$, the
contribution to the potential of the small antidot at the center
of the unit cell is strongly repulsive. When additional vortices enter the sample, they will be pushed
towards the large antidots, leading to an increase of their
effective saturation number. In other words,
{\it the additional repulsive potential at the small antidots helps to
increase the saturation number of the larger antidots}. Within this scenario, for $H>H_2$, the phase boundary \tch~ of the
composite antidot lattice with a saturated smaller antidot
resembles strongly the phase boundary expected for a simple square
antidot lattice, without small antidots in the center, but with a
larger antidot size. Due to this larger effective size, these
antidots are then able to trap more vortices. \textit{We
therefore conclude that of the four flux quanta trapped per unit
cell of the composite antidot lattice, one is pinned by the small
antidot, while three are pushed into the larger holes.} The net
effect of the addition of the smaller hole in the antidot film, is
to increase the effective pinning capacity (or effective
saturation number) of the lattice with large antidots. This leads
to a substantial broadening of the field range where a strong
\tch~ enhancement is observed. A similar picture was introduced by Doria and co-workers to explain the multiple trapping of vortices at high fields, as a result of the pressure exerted by the external vortices into the pinning site.\cite{doria}

\begin{figure}[hctb]
  \centering
  \caption{Schematic representation of the potential along a diagonal of the
  composite antidot array (see inset), experienced by a vortex entering the sample for $H=0$, $H_1$, and $H_2$.}
  \label{fig:ChC_potential}
\end{figure}

The potential at $H=H_2$, drawn schematically in the lower panel of
Fig.~\ref{fig:ChC_potential}, can be seen as a checkerboard
pattern with consecutively a repulsive and an attractive site. In
a recent calculation, Lindquist and Riklund have modified the
classical problem of a two-dimensional electron gas exposed to a
magnetic field \cite{hofstadter} by adding a periodic
checkerboard-like on-site potential with alternating signs
\cite{lindquist}. Since the lowest energy level $E_{LL}(H)$,
found for this classical problem, corresponds to the \tch~ phase
boundary of a simple square antidot lattice in the ``collective''
regime, we believe that
the addition of the checkerboard on-site potential in the theory
corresponds to our experimental situation for the composite
antidot lattice at fields higher than the second matching field
($H > H_2$). These authors showed that both the integer and the rational
matching features in $E_{LL}(H)$ are smeared out in the checkerboard
system. These results may explain why the matching features in
\tch~ are much weaker for $H > H_2$, than for $H < H_2$ (see Fig.~\ref{fig:ChC_tchfit}).
Indeed, it is only when $H_2$ is exceeded that the
checkerboard-like pinning potential is realized experimentally.

Summarizing the results of the \tch~ phase boundary measurements,
the composite antidot lattice shows enhanced pinning, with many
integer matching features appearing for fields up to $H_6$. The
broadening of the \rt~ transition after $H_4$ indicates that at
least four flux quanta can be trapped per antidot. From the shift
in magnetic field of the parabolic background of \tch~ with one
matching period, we have deduced that the small antidot pins a
$\Phi_0$-vortex, while $3 \po$-vortices are trapped by the
large antidots. The presence of the additional small antidot in
the center of the unit cell has therefore led to a substantial
broadening of the collective regime, or, in other words, to an
increase of the effective saturation number $n_s$ of the large
antidots from one to three.

\section{Critical current as a function of field}
\label{section:ChC_ich}

So far we have explored the normal-superconducting boundary in order to experimentally determine the extension of the different regimes given by the ratio $\xi(T)/(d-a)$. Now we turn to isothermal critical current measurements which allow us to study the vortex dynamics deep in the superconducting state.

The critical current as a function of field \ich~ was measured using a 100 $\mu$V voltage criterion for several temperatures close to \tch. The results, in order of
decreasing temperature, are shown in Figs.~\ref{fig:ChC_ich0997_3}
and \ref{fig:ChC_ich0986_74}.

The absolute value of the critical current density at zero field
for the composite antidot array amounts to $I_{co}= I_{c}(H=0) = 6.8~\cdot~10^{8}
\frac{\mathrm{A}}{\mathrm{m}^2}$ at $T/\tco=0.974$. 
This value is a factor $\sim 3$ lower than the critical current density obtained for the reference antidot lattice, in part due to the difference in $\lambda(0)$ and $\xi(0)$, which eventually determine the pinning properties. 
The specific geometry of the lateral nanopatterning might also influence the current distribution throughout the film, hereby also affecting the critical current $I_{co}$. In order to compare the pinning properties of the film with the composite antidot lattice (open symbols) and the reference antidot lattice (solid lines) measured at the same reduced temperature we have normalized the critical current by $I_{co}$. Notice that since the saturation number $n_s$ is mainly determined by the coherence length\cite{schmidt,nordborg} $\xi(T)$ which in turn depends solely on the reduced temperature $t=T/T_c$, {\it regardless the value of the applied magnetic field}, it is enough to compare the results obtained on these samples at the same reduced temperature, without the necessity of normalizing the field.

\begin{figure}[hctb]
  \centering
  \caption{Normalized critical current at
  $T/T_{c0}=0.997$ and $T/T_{c0}=0.993$ of a film with a
  composite antidot array. The curves were measured for $H<0$
  (open symbols) and symmetrized for clarity for $H>0$ (dashed line).
  For comparison, the solid line shows the normalized critical current
  \ich/$I_{c0}$ obtained for the film with the reference antidot lattice.}
  \label{fig:ChC_ich0997_3}
\end{figure}

\begin{figure}[hctb]
  \centering
  \caption{Normalized critical current at
  $T/T_{c0}=0.986$ and $T/T_{c0}=0.974$ of a film with a
  composite antidot array. The curves were measured for $H<0$
  (open symbols) and symmetrized for clarity for $H>0$ (dashed line).
  For comparison, the solid line shows the normalized critical current
  \ich/$I_{c0}$ obtained for the film with the reference antidot lattice.}
  \label{fig:ChC_ich0986_74}
\end{figure}

The  \ich/$I_{c0}$ curves for the film with a composite antidot array
(Figs.~\ref{fig:ChC_ich0997_3} and \ref{fig:ChC_ich0986_74}) have
been measured for negative fields (open symbols) and
symmetrized for $H>0$ for clarity (dashed lines). All curves show
distinct periodic matching features, with a period
$H_1=\frac{\po}{d^2}=9.2~\mathrm{G}$ corresponding to the unit
cell of the lattice with the large (or the small) antidots
($d=1.5~\mu$m). As we pointed out before, the
periodicity felt by the vortices is defined by one of the
interpenetrating antidot lattices of the composite array, and not
by the resulting square lattice with a unit cell which is twice as
small and rotated by 45$^\circ$.

In the upper panel of Fig.~\ref{fig:ChC_ich0997_3} ($T/\tco=0.997$),
the  \ich/$I_{c0}$ curve of the film with a composite antidot lattice
shows a peak-like behavior with sharp maxima at $H_1$, $H_2$, and
$H_3$. This behavior is expected at temperatures sufficiently
close to $\tco$, where it is not possible to have interstitial
vortices in the superconducting strands between the antidots. As
we have already deduced from the shape of the \tch~ phase
boundary (Section~\ref{composite_tch}), interstitial vortices
indeed appear in the sample only for $T/\tco \leq 0.994$.

At a lower temperature, $T/\tco=0.993$
(Fig.~\ref{fig:ChC_ich0997_3}, lower panel), and all temperatures
below that (Fig.~\ref{fig:ChC_ich0986_74}), a strong enhancement
of  \ich/$I_{c0}$ in the film with a composite antidot lattice can be
found for fields higher than the first matching field $H_1$,
compared to the reference antidot lattice. The reason for this
lies in the \textit{ability of the composite antidot lattice to
pin more flux quanta inside the antidots compared to the reference
antidot array} (see Section~\ref{composite_tch}).

It should be noted that the field range where the film has a finite critical current, i.e.
where the film remains superconducting, is considerably broader
for the composite than for the
reference antidot array. 

The appearance and sharpness of the matching features in the
\ich/$I_{c0}$ curves, are temperature
dependent. At $T/\tco=0.993$ (Fig.~\ref{fig:ChC_ich0997_3}, lower
panel), every integer matching peak up to $H_6$ can be clearly
seen. The maxima at $H_1$, $H_2$, and $H_3$ are very pronounced.
At $H_4$ and $H_5$, one finds cusps rather than local maxima in
 \ich/$I_{c0}$. The matching feature at $H_6$ is again peak-like. This
indicates that the vortex patterns formed at $H_4$ and $H_5$ are
less stable than the vortex configuration at $H_6$.

When the temperature is lowered to $T/\tco=0.986$
(Fig.~\ref{fig:ChC_ich0986_74}, upper panel), we find again sharp
matching features in  \ich/$I_{c0}$ at $H_1$, $H_2$ and $H_3$, and only
very weak cusps at $H_4$ and $H_5$. At $H_6$, the local maximum
has developed into a pronounced cusp, after which a substantial
change in the  \ich/$I_{c0}$ slope occurs. A second smaller slope change
can be found at $H_7$. At the lowest measured temperature,
$T/\tco=0.974$ (Fig.~\ref{fig:ChC_ich0986_74}, lower panel), the
only matching features left are the sharp local maxima at $H_1$,
$H_2$, and $H_3$, and one pronounced cusp at $H_7$. It appears
that, at this temperature, the seventh matching field $H_7$ plays
the same role as the sixth matching field $H_6$ at $T/\tco=0.986$.
This fact leads us to believe that at $T/\tco=0.974$, \textit{the
total number of trapped flux quanta per unit cell of the composite
lattice, has increased from four to five}.

It is worth to notice that the normalized critical current at the first matching field $H_1$ reaches
approximately the same value for the film with the composite and
with the reference antidot lattice, except for the \ich/$I_{c0}$ curve
taken at $T/\tco=0.997$. This fact makes the film with the reference antidot array
a good candidate to compare its pinning properties with
those of the composite antidot array.

\section{Stable vortex patterns}
\label{patterns}

The periodic matching features in the \tch~ phase boundary and in
the critical current versus field curves \ich~ demonstrate that
the composite antidot lattice can stabilize commensurate vortex
lattices at several magnetic fields. From the results presented in
Section~\ref{composite_tch}, it is clear that the large antidots
trap at least three flux quanta, while the smaller antidots are
saturated after pinning one single quantum vortex. The vortex
patterns expected in the composite antidot lattice will therefore
differ from the known patterns in simple square
pinning arrays.

\begin{figure}[tb]
  \centering
  \caption{Suggested vortex pattern at $H_4$, $H_5$, $H_6$, and $H_7$.
  All patterns have been obtained by molecular dynamics simulations by an
  annealing procedure, except the one at $H_4$. Open circles and black
  dots represent pinning sites and single quantum vortices, respectively.}
  \label{fig:ChC_patternh6}
\end{figure}

We have performed molecular dynamics simulations to obtain the
vortex patterns at the matching fields $H_5$, $H_6$, and $H_7$. To model the composite vortex lattice, two interpenetrating arrays
of Gaussian sites with a different radius and a different pinning
force were used. This was necessary to obtain an occupation of $3
\po$-vortices in the large pinning sites, and of one $\po$-vortex
in the small pinning sites. By applying an annealing course, in
which the temperature is lowered, starting from a high temperature
and a random distribution of vortices, the most stable
configuration in the given pinning potential can be found. As the
temperature is lowered, the pinning sites become smaller and
stronger, scaling with $\xi(T)$ in the following way
\cite{zhu}:

\begin{equation}
F_p \propto F_{p0} \exp \left(- \frac{r}{\xi(T)} \right)^2 \, ,
\label{eq:simulationpotential}
\end{equation}
where $F_{p0}$ is the single site pinning force. Additionally, the
vortex-vortex interaction range reduces with decreasing
temperature, due to the decrease of the magnetic penetration depth
$\lambda$. In this type of
simulation, the occupation of the pinning sites lowers with
decreasing temperature. The annealing course was therefore stopped
when the occupation, corresponding to our experimental situation,
was achieved. Figure~\ref{fig:ChC_patternh6} shows the vortex
configurations we suggest for $H_4$, $H_5$, $H_6$, and $H_7$.  All
patterns, except the one at $H_4$, are obtained from the molecular
dynamics simulations. Multi-quanta vortices are represented in
this model by a multiple occupation of a pinning site with
(repulsive) single-quantum vortices. Since in the experiment, the
pinning sites consist of real holes in the film, the
vortices trapped in the same pinning site will be interpreted as
multi-quanta vortices, even though they are depicted as separate
single flux quantum entities in the plots. Actually, this model can be experimentally realised by an array of non fully perforated (or blind) holes. Reported results on such systems showed that blind holes are weaker pinning centers than antidots, although the overall features in both cases are very much alike.\cite{blind-holes} 

The vortex pattern at $H_4$, which is drawn schematically and was
not calculated, depicts all antidots occupied with the maximum
number of vortices. The large antidots trap $3\po$-vortices, the
smaller antidots trap a $\po$-vortex. No interstitial vortices are
present in the sample. In this case, one would expect the
matching feature at $H_4$ to be of the same kind as the one at
$H_3$. Surprisingly, the \ich~ curves (see e.g.
Fig.~\ref{fig:ChC_ich0986_74}) show only very weak matching
features at $H_4$. However, the \tch~ measurements and the fact
that the matching peak in \ich~ at $H_6$ is very well defined
(see discussion below), both indicate that there should be four vortices
trapped per unit cell of the antidot array, leading to the
suggested vortex pattern.

At $H_5$, there is one interstitial vortex present per unit cell
of the array. It is, however, not evident where this
vortex is located, since the most logical position, at the center of the unit
cell, is already occupied by the filled smaller antidot. One can
see a tendency of the interstitial vortices to form diagonal
lines, which make zigzag traces across the sample, indicated by
dashed lines in Fig.~\ref{fig:ChC_patternh6}(b). However, the long
range order which is needed to make a regular pattern, with for
example straight diagonal lines or a regular zigzag, is lacking at
this field. The pattern found in molecular dynamics simulations
for $H_5$ is consequently not very stable.

At $H_6$, a highly symmetric vortex pattern is formed. In
this case, two interstitial vortices are present per unit cell,
which are positioned approximately at the center of the line
connecting two neighboring large antidots. Due to its high
symmetry, the vortex pattern at $H_6$ is very stable. Remarkably,
the calculations for $H_6$, resulting in a very regular pattern,
have been performed under the same conditions as the ones at
$H_5$, where no regular pattern could be found. This is an
indication of the different stability of the vortex patterns at
$H_5$ and $H_6$. Indeed, the matching feature at $H_6$ in \tch~
or  \ich/$I_{c0}$ is always more pronounced than at $H_5$ (see e.g.
Figs.~\ref{fig:ChC_deltatch} and \ref{fig:ChC_ich0986_74}, upper
panel). For the  \ich/$I_{c0}$ curve measured at $T/\tco=0.974$
(Fig.~\ref{fig:ChC_ich0986_74}, lower curve), the matching cusp at
$H_7$ becomes rather sharp. We suggest that at this temperature,
the large antidots are able to trap four flux quanta instead of
three. In that case, the expected vortex pattern for $H_7$ (see
Fig.~\ref{fig:ChC_patternh6}), resembles the pattern
calculated for the sixth integer matching field, but
with four flux quanta occupying the large antidots instead of
three.

To obtain a regular pattern at $H_7$, with three flux quanta
pinned in the large antidots, three interstitial vortices have to
be accommodated per unit cell of the composite array. The
calculations were not able to produce a regular vortex pattern
with an occupation of three at the large pinning sites and one
at the smaller pinning sites.  This indicates that the stability
of a vortex pattern at $H_7$ is not very high. On the other hand,
the calculation method, where $3~\po$-vortices are represented by
three separate $\po$-vortices, might in this case also affect the
outcome of the simulation. Indeed, the cylindrical symmetry of the
$3~\po$-vortex, as it occurs in the experiment, will make it
easier to obtain a regular vortex pattern.

The vortex patterns suggested in this section remain to be directly verified by using a local scanning technique, such as low
temperature scanning Hall probe microscopy. Further insight into
the vortex pinning and dynamics in systems with a composite
pinning array may also be gained from molecular dynamics
simulations.

\section{Conclusions}

We have used a composite antidot lattice, consisting of two
interpenetrating antidot arrays with a different antidot size, but
with the same lattice period, as a strong periodic pinning
potential for the vortex lattice in a superconducting film. The
shift between the two lattices is such that the smaller antidots
are situated exactly at the centers of the cells of the array
of large antidots. We have shown that this pinning array can
stabilize the vortex lattice at several matching fields from $H_1$
to $H_7$.

Measurements of the critical temperature \tch~ and current
 \ich/$I_{c0}$ as a function of magnetic field, have demonstrated that
the composite antidot lattice can trap a considerably higher
amount of flux quanta per unit cell (four or five instead of one)
inside the antidots, compared to a reference antidot film without
the additional small antidots in the center of the unit cell. This
means that the appearance of interstitial vortices in the
composite antidot lattice is delayed to higher magnetic fields.
The presence of the smaller antidots has therefore increased the
effective saturation number of the large antidots, which has led
to a considerable expansion of the field range in which an
enhanced critical current is observed.

\acknowledgments
This work was supported by the Research Fund K.U.Leuven GOA/2004/02, the Belgian Interuniversity Attraction Poles (IUAP), the Fund for Scientific Research Flanders (FWO) and ESF ``VORTEX'' program.

\bibliographystyle{prsty}

\begin{thebibliography}{10}

\bibitem{moshchalkov}
V. V. Moshchalkov, M. Baert, E. Rosseel, V. V. Metlushko, M. J. Van Bael, and Y. Bruynseraede, Physica C {\bf 282}, 379 (1997). V. V. Moshchalkov, M. Baert, V. V. Metlushko, E. Rosseel, M. J. Van Bael, K. Temst, and Y. Bruynseraede, Phys. Rev. B {\bf 57}, 3615 (1998). 

\bibitem{metlushko}
V. Metlushko, U. Welp, G. W. Crabtree, R. Osgood, S. D. Bader, L. E. DeLong, Zhao Zhang, S. R. J. Brueck, B. Ilic, K. Chung, and P. J. Hesketh, Phys. Rev. B {\bf 60}, R12585 (1999).

\bibitem{welp}
U. Welp, Z. L. Xiao, J. S. Jiang, V. K. Vlasko-Vlasov, S. D. Bader, G. W. Crabtree, J. Liang, H. Chik, and J. M. Xu, Phys. Rev. B {\bf 66}, 212507 (2002).

\bibitem{rosseel}
E. Rosseel, T. Puig, M. Baert, M. J. Van Bael, V. V. Moshchalkov, and Y. Bruynseraede, Physica C {\bf 282}, 1567 (1997).

\bibitem{sophie-crete}
S. Raedts, A. V. Silhanek, M. J. Van Bael, and V. V. Moshchalkov, Physica C {\bf 404}, 298 (2004).

\bibitem{handbookchemistryandphysics}
R. C. Weast (ed.), Handbook of Chemistry and Physics (the Chemical Rubber Co., Ohio, 1971).

\bibitem{lambda}
T. P. Orlando and K. A. Delin, "Foundations of Applied Superconductivity," Addison-Wesley, Reading, MA (1991).

\bibitem{doria}
M. M. Doria and G. F. Zebende, Phys. Rev. B {\bf 66}, 064519 (2002).

\bibitem{wahl}
A. Wahl, V. Hardy, J. Provost, Ch. Simon, A. Buzdin, Physica C {\bf 250}, 163 (1995).

\bibitem{comment}
Additionally to the perforation effect, the finite thickness $\delta$ of the film also yields an increase of the effective penetration depth up to $\Lambda=2\lambda^2/\delta$ therefore giving rise to an even higher $\kappa$ value. This issue has been briefly addressed in V. V. Moshchalkov, M. Baert, V. V. Metlushko, E. Rosseel, M. J. Van Bael, K. Temst, R. Jonckheere, and Y. Bruynseraede, Phys. Rev. B {\bf 54}, 7385 (1996).

\bibitem{pannetier}
B. Pannetier in `` Quantum Coherence in Mesoscopic Systems", edited by B. Kramer (Plenum Press, New York, 1991), chapter 9, pp. 457-484.

\bibitem{hofstadter}
D. R. Hofstadter, Phys. Rev. B {\bf 14}, 2239 (1976).

\bibitem{lindquist}
B. Lindquist and R. Riklund, Phys. Rev. B {\bf 60}, 10054 (1999).

\bibitem{schmidt}
G.S. Mkrtchyan and V.V. Schmidt, Sov. Phys. JETP {\bf 34}, 195 (1972).

\bibitem{nordborg}
H. Nordborg and V. M. Vinokur, Phys. Rev. B {\bf 62}, 12408 (2000).

\bibitem{zhu}
B. Y. Zhu, L. Van Look, and V. V. Moshchalkov, B. R. Zhao, and Z. X. Zhao, Phys. Rev. B {\bf 64}, 012504 (2001).

\bibitem{blind-holes}
S. Raedts, A.V. Silhanek, M. J. Van Bael, and V.V. Moshchalkov, Phys. Rev. B {\bf 70}, 024509 (2004).

\end{thebibliography}

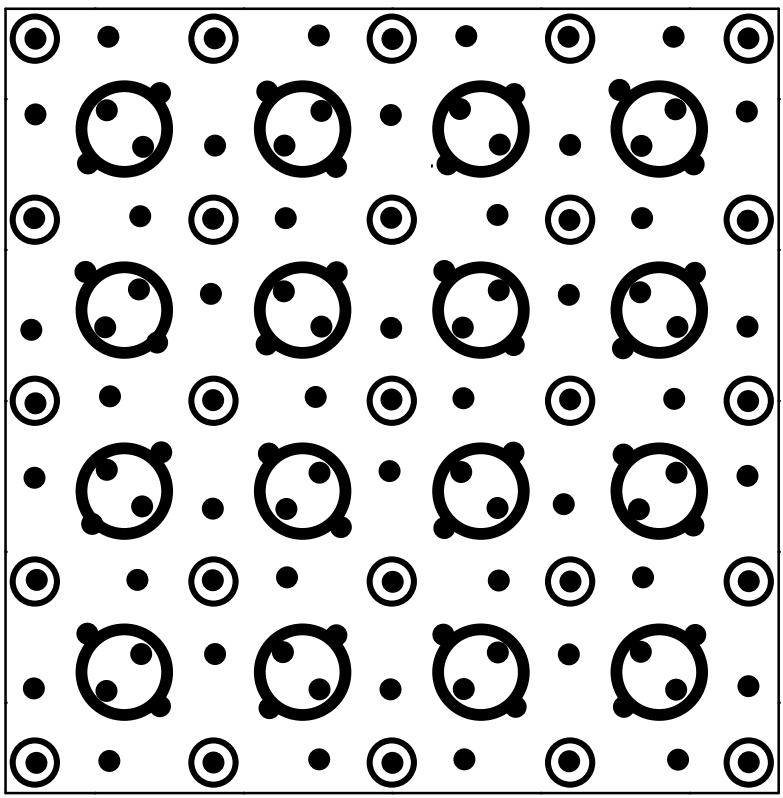
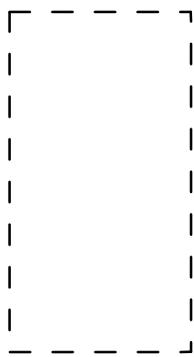
\end{document}